\documentclass[a4paper]{spie}  
\usepackage{amsmath,amsfonts,amssymb}

\usepackage[utf8]{inputenc}   
\usepackage[english]{babel}   
\usepackage[usenames,dvipsnames]{xcolor} 
\usepackage[colorlinks=true, allcolors=blue, pdftex]{hyperref}
\usepackage{aas_macros}
\usepackage{paralist}
\usepackage{graphicx}

\title{Architecture of the SOXS instrument control software}

\author[*,a]{Davide Ricci}
\author[a]{Andrea Baruffolo}
\author[a]{Bernardo Salasnich}
\author[a]{Daniela Fantinel}
\author[b]{Josefina Urrutia}

\author[c]{Sergio Campana}
\author[a]{Riccardo Claudi}
\author[d]{Pietro Schipani}
\author[c]{Matteo Aliverti}
\author[e]{Sagi Ben-Ami}
\author[a]{Federico Biondi}
\author[b]{Anna Brucalassi}
\author[d]{Giulio Capasso}
\author[f]{Rosario Cosentino}
\author[g]{Francesco D'Alessio}
\author[c]{Paolo D'Avanzo}
\author[h]{Oz Diner}
\author[j,q]{Hanindyo Kuncarayakti}
\author[k]{Matteo Munari}
\author[h]{Adam Rubin}
\author[k]{Salvatore Scuderi}
\author[g]{Fabrizio Vitali}
\author[l]{Jani Achrén}
\author[m,t]{José Antonio Araiza-Duran}
\author[n]{Iair Arcavi}
\author[c]{Andrea Bianco}
\author[a]{Enrico Cappellaro}
\author[d]{Mirko Colapietro}
\author[d]{Massimo Della Valle}
\author[d]{Sergio D'Orsi}
\author[o]{Johan Fynbo}
\author[h]{Avishay Gal-Yam}
\author[c]{Matteo Genoni}
\author[p]{Mika Hirvonen}
\author[j,q]{Jari Kotilainen}
\author[q]{Tarun Kumar}
\author[c]{Marco Landoni}
\author[p]{Jussi Lehti}
\author[r]{Gianluca Li Causi}
\author[a]{Luca Marafatto}
\author[q]{Seppo Mattila}
\author[c]{Giorgio Pariani}
\author[m,t]{Giuliano Pignata}
\author[h]{Michael Rappaport}
\author[c]{Marco Riva}
\author[s]{Stephen Smartt}
\author[a]{Massimo Turatto}
\author[k]{Ricardo Sanchez}

\affil[a]{INAF -- Osservatorio Astronomico di Padova, Vicolo dell’Osservatorio 5, I-35122, Padua, Italy }
\affil[b]{ESO, Karl Schwarzschild Strasse 2, D-85748, Garching bei München, Germany }
\affil[c]{INAF -- Osservatorio Astronomico di Brera, Via Bianchi 46, I-23807, Merate, Italy }
\affil[d]{INAF -- Osservatorio Astronomico di Capodimonte, Sal. Moiariello 16, I-80131, Naples, Italy }
\affil[e]{Harvard-Smithsonian Center for Astrophysics, Cambridge, USA }
\affil[f]{FGG-INAF, TNG, Rambla J.A. Fernández Pérez 7, E-38712 Breña Baja (TF), Spain }
\affil[g]{INAF -- Osservatorio Astronomico di Roma, Via Frascati 33, I-00078 M. Porzio Catone, Italy }
\affil[h]{Weizmann Institute of Science, Herzl St 234, Rehovot, 7610001, Israel }
\affil[j]{Finnish Centre for Astronomy with ESO (FINCA), FI-20014 University of Turku, Finland}
\affil[k]{INAF -- Osservatorio Astrofisico di Catania, Via S. Sofia 78 30, I-95123 Catania, Italy }
\affil[l]{Incident Angle Oy, Capsiankatu 4 A 29, FI-20320 Turku, Finland }
\affil[m]{Universidad Andres Bello, Avda. Republica 252, Santiago, Chile }
\affil[n]{Tel Aviv University, Department of Astrophysics, 69978 Tel Aviv, Israel }
\affil[o]{Dark Cosmology Centre, Juliane Maries Vej 30, DK-2100 Copenhagen, Denmark }
\affil[p]{Aboa Space Research Oy, Tierankatu 4B, FI-20520 Turku, Finland}
\affil[q]{Tuorla Observatory, Dept. of Physics and Astronomy, FI-20014 University of Turku, Finland }
\affil[r]{INAF - Istituto di Astrofisica e Planetologia Spaziali, Rome, Italy}
\affil[s]{Astrophysics Research Centre, Queen's University Belfast, Belfast, BT7 1NN, UK }
\affil[t]{Millennium Institute of Astrophysics (MAS)}

\authorinfo{$^*$Contact information:
  D.R: davide.ricci@inaf.it, +39-049-829-3480
}

\begin{document}
\maketitle

\begin{abstract}
  SOXS (Son Of X-Shooter) is a new spectrograph for the ESO NTT
  telescope, currently in the final design phase.
  The main instrument goal is to allow the characterization of
  transient sources based on alerts.  It will cover from near-infrared
  to visible bands with a spectral resolution of $R\sim 4500$ using
  two separate, wavelength-optimized spectrographs.  A visible camera,
  primarily intended for target acquisition and secondary guiding,
  will also provide a scientific ``light'' imaging mode.
  In this paper we present the current status of the design of the
  SOXS instrument control software, which is in charge of controlling
  all instrument functions and detectors, coordinating the
  execution of exposures, and implementing all observation,
  calibration and maintenance procedures.
  Given the extensive experience of the SOXS consortium in the
  development of instruments for the VLT, we decided to base the
  design of the Control System on the same standards, both for
  hardware and software control.
  We illustrate the control network, the instrument functions and
  detectors to be controlled, the overall design of SOXS Instrument
  Software (INS) and its main components.  Then, we provide details
  about the control software for the most SOXS-specific features:
  control of the COTS-based imaging camera, the flexures compensation
  system and secondary guiding.
\end{abstract}

\keywords{SOXS, Instrument Control Software, Software, Spectroscopy,
  Imaging, Astronomy}

\section{Introduction}
\label{sec:intro}

SOXS, which stands for ``Son Of X-Shooter'', is a new
instrument\cite{2016SPIE.9908E..41S} being developed for the European
Southern Observatory (ESO) New Technologies Telescope (NTT) at the La
Silla Observatory, Chile, inspired to the X-Shooter
spectrograph\cite{2011A&A...536A.105V} at the Very Large Telescope
(VLT).
This new, transient-oriented facility will be mainly dedicated to
several follow-up programs for the characterization of sources based
on alerts.  These will come from ``traditional'' telescope surveys,
from high-energy to radio astronomy, and will span to multi-messenger
Astronomy, with the inclusion of neutrinos and gravitational wave
experiments \cite{2017arXiv171005915B}.
The instrument covers from near-infrared to visible bands (350 -- 2000
nm) and has a spectral resolution of $R\sim 4500$. It features two
spectrographs: the first one optimized for the near-infrared
wavelengths \cite{soxsvitali}, while the second one optimized for the
visible wavelength range \cite{soxscosentino}.  An imaging camera provides a
scientific imaging mode \cite{soxsbrucalassi}, as well as support for
target acquisition and secondary guiding capabilities.

\begin{figure} [tbp]
  \centering
  \includegraphics[width=0.71\textwidth]{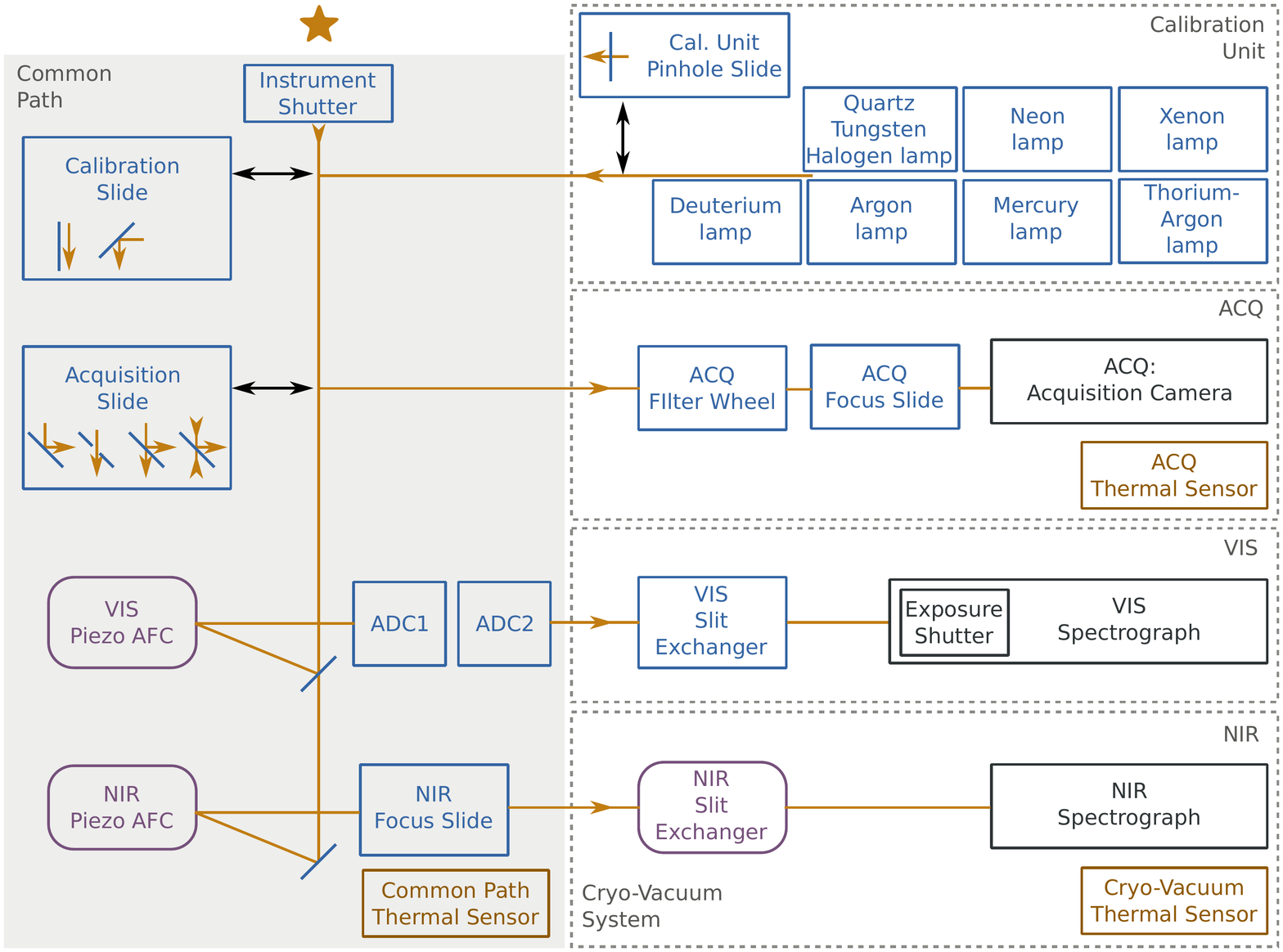}
  \caption[schematic-diagram]
  { \label{fig:schematic-diagram} The SOXS instrument from the
    Instrumentation Software point of view.  Boxes represent
    instrument devices, sensors and detectors. Rounded boxes represent
    non-VLT standard devices. The path of the light is also indicated for
    clarity. }
\end{figure}

SOXS is composed by several sub-systems, shown in
Fig.~\ref{fig:schematic-diagram} which here we briefly describe:
\begin{description}
\item[Common Path (CP):] is in charge for relaying the light from the
  NTT focal plane to the entrance of the spectrographs. In doing that,
  the CP selects the wavelength range for the spectrographs (using a
  dichroic) and changes the focal ratio.  A number of devices are
  present in the CP and have to be controlled by the Instrument
  Software (INS).
  The first is the entrance shutter, used to let the telescope beam in
  or make the instrument light-tight when performing calibrations
  using internal light sources.
  A linear motor allows to select the instrument input source:
  either the lamps of the Calibration Unit or the light coming from
  the target field on sky.
  Then, a motorized slide allows to direct all the beam to the
  Acquisition Camera, for Imaging observation, or to transmit the
  central part of the beam to the spectrographs, while the periphery
  is sent to the Acquisition Camera for secondary guiding.  Two
  additional positions are used for calibrations and maintenance.
  Two separate piezo-electric Tip-Tilt Mirrors are used to compensate
  for mechanical flexures due to the changing of the gravity vector
  during observations.
  The CP also provides an Atmospheric Dispersion Corrector (ADC) for
  the visible spectrograph, which include two motors, and a linear
  stage providing adjustment of the focus for the near-infrared
  spectrograph.

\item[Calibration Unit (CU):] includes two sets of lamps for
  wavelength and flat-field calibration, and an insertable pinhole for
  alignement.

\item[Acquisition Camera (ACQ):] allows to center the source on the
  selected slit during target acquisition and to perform secondary
  guiding during observations. It can be used also as light imager in
  order to perform photometry and flux calibration.  It is composed by
  a commercial \emph{Andor iKon-M 934 Series} Camera, equipped with a
  deep-depletion CCD and providing a $3.5^\prime\times 3.5^\prime$
  Field of View (FoV).
  ACQ also includes a linear motor for focusing, and an eight
  positions filter wheel.

\item[Visible spectrograph (VIS):] it is based
  on a new design in which the spectral band is split in narrow
  sub-bands, thus allowing the use of high efficiency gratings each
  optimized for a narrow wavelength range.
  Besides the fixed optics, it comprises an \emph{E2V CCD44-82}
  $2 \rm k \times 4 \rm k$ CCD detector, its front-end electronics,
  the mechanical support and a five position slit exchanger.

\item[Near-infrared spectrograph (NIR):] this sub-unit is composed by
  an echelle-dispersed spectrograph working in the 800--2000 nm
  wavelength range. It is enclosed in a cryostat supplemented by a
  Cryo-Vacuum controller.
  Its detector system comprises a \emph{Teledyne H2RG TM}
  $2048 \times 2048$ pixels hybrid infrared array detector, its
  front-end electronics and controller. The mechanical support, and a
  cryogenic, piezo-mechanic slit exchanger, with the same options of
  the VIS, complete this subsystem.

\item[Cryo-Vacuum Control Sub-system (CVS):] consists of all the
  hardware and the electronics related to control and monitoring of
  the cryogenic functions for the two cryostats (enclosing the UV-VIS
  detector and the whole NIR spectrograph, respectively).
\end{description}

The SOXS project went through the Preliminary Design Review (PDR) in
July 2017 and is currently approaching the Final Design Review (FDR),
scheduled for July 2018.
In this paper we present the current status of the Instrument Software
(INS).  This paper is part of a series of contributions
\cite{soxsvitali,soxscosentino,soxsbrucalassi,
  soxsaliverti,soxsbiondi,soxscapasso,soxsclaudi,soxssanchez,soxsschipani,soxsrubin}
describing the SOXS design and properties after the instrument PDR.

The control network architecture is presented in Sect.~\ref{sec:net},
and the overall design of the Instrument Software is described in
Sect.~\ref{sec:sof}.  Then, we give some more details about the design
of the ACQ camera software (Sect.~\ref{sec:cam}), the flexure control
system (Sect.~\ref{sec:afc}), and the secondary guiding software
(Sect.~\ref{sec:sg}).  The development status is treated in
Sect.~\ref{sec:dev}, and conclusions are presented in
Sect.~\ref{sec:conc}.

\section{Network architecture}
\label{sec:net}

\begin{figure} [tbp]
  \centering
  \begin{tabular}{cc}
    \includegraphics[width=0.50\textwidth]{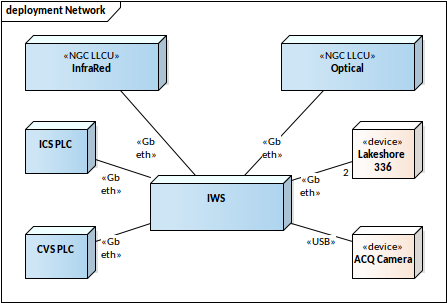}
    \includegraphics[width=0.48\textwidth]{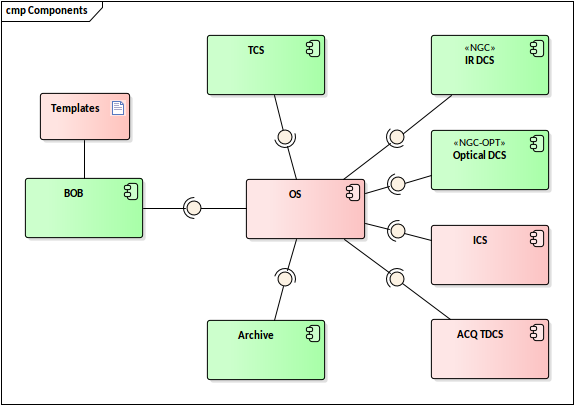}
  \end{tabular}
  \caption[control-network]
  { \label{fig:control-network}
    Left: Network control architecture of SOXS.
    Right: Components of the SOXS software;
    red boxes represent software to be configured or developed, while
    green boxes represent VLTSW components that will be used without
    modifications.}
\end{figure}

The control network of SOXS is illustrated in the deployment diagram
in the left panel of Fig.~\ref{fig:control-network}. The architecture
is the typical one for control system of VLT Instruments: a number of
local controllers are connected to a supervising Instrument
Workstation (IWS) through an Instrument LAN, based on Gb Ethernet.
Following the recently-introduced VLT
standard\cite{2014SPIE.9152E..07K}, all instrument functions are
controlled by a single \emph{Beckhoff} Programmable Logic Controller
(PLC).
A separate \emph{Siemens} S7 PLC is responsible for the CVS functions
and sensors control.  The CVS is an autonomous system, SOXS INS is not
in charge of controlling but only monitoring it.

The detector controllers for the VIS and NIR channels are instances of
the ESO New General Detector Controller (NGC). The respective Linux
Local Control Units (LLCUs) are connected to the IWS via the Instrument
LAN.

%
On the other hand, the ACQ camera is based on a Commercial, Off-The
Shelf (COTS) component, providing an integrated controller and a USB
interface.  In the baseline design it is foreseen to route the USB
connection to the IWS through the Observatory LAN by means of a
commercial USB extender.

Finally, NIR and VIS detectors temperatures are controlled by means of
two \emph{Lakeshore} 336 controllers, which are directly connected to
the IWS through Gigabit Ethernet.
%

\section{Software architecture}
\label{sec:sof}

The SOXS Instrument Software is in charge of the control of all
instrument functions and monitoring of all sensors. It shall also
manage external interfaces: with the High-level Observation Software
(HOS), the Telescope Control Software (TCS), and the Archive.  INS
also implements all procedures relative to observation, calibration
and maintenance operations, in the form of templates.

SOXS in an instrument for the NTT, and therefore it is not mandatory
that it complies with VLT standards. However, given the large
experience in our group in the development of control software for VLT
instruments, it has been decided to adopt the VLT standards also for
SOXS INS.

Therefore, the SOXS INS architecture follows the standard partitioning
of VLT Instrument Software applications.  All instrument functions
except detectors are controlled by an Instrument Control Software
(ICS), science detectors are controlled by instances of Detector
Control Software (DCS), which, in this case, are based on NGC
software.  On the other hand, since the Andor camera is not a standard
VLT component, its control software is an application based on the
Technical DCS Software Development Kit (SDK) which is part of the
VLTSW2016 release we are using\cite{2014SPIE.9152E..0ID}.

The ICS provided by VLTSW supports a number of ``standard devices'',
like linear stages, shutters, lamps, etc., for which no code
development is necessary, but it is enough to provide configuration
information (location and type of signals, motor control parameters,
etc.) For non-standard devices, some development is required to
properly interface them with ICS. In SOXS control system, two devices
are non-standard: the piezo-actuated tip-tilt mirrors used for flexure
compensation and the cryogenic piezo-mechanic stage used for slit
positioning in the NIR spectrograph.

On the other hand, since the Andor camera is not a standard VLT
component, its control software is an application based on the
Technical DCS Software Development Kit (SDK) which is part of the
VLTSW2016 release we are using\cite{2014SPIE.9152E..0ID}.

Operations are coordinated by a central component, the Observation
Software (OS), which is also in charge of managing external
interfaces. All observation, calibration and maintenance procedures
are implemented as templates, which are executed by the Broker of
Observation Blocks (BOB) through commands sent to the OS.

These SOXS INS components are displayed in the right panel of
Fig.~\ref{fig:control-network}.

As already mentioned, the ACQ camera and the Active Flexure
Compensation system are special components of the SOXS instrument
(i.e. not directly supported by the VLT Instrument Software
framework). Therefore we briefly describe them in the following subsections.

\subsection{Acquisition Camera control}
\label{sec:cam}

The imaging and secondary guiding camera of SOXS is a Commercial
Off-the-Shelf (COTS) component, provided with a USB 2.0 interface for
external connection.
Since the SOXS IWS will be located about 3 km from the ACQ, while
USB cable connections are limited to a few meters, we decided to use a
commercial, switchable USB extender (\emph{Icron
  2304GE-LAN}\cite{IcronLink}) to route the connection through the La
Silla LAN and then to the IWS.  This solution, however, has not been
validated on the field, and tests are underway at the Observatory to
verify its feasibility.

Therefore, a fall-back design is being considered in which an
intermediate computer will be located aboard the instrument, close to
the camera and directly connected to it via USB, and will act as a
``software gateway'' towards the IWS.

In either case, the ACQ camera control will be based on
TDCS\cite{2014SPIE.9152E..0ID} SDK, and its core will consists in a
class providing the ``communication interface'' with the camera.  The
main difference will be that in the baseline, this class will directly
interface with the Andor camera via USB, directly calling the
vendor-supplied driver functions, while in the fall-back design, the
communication will happen through the Instrument LAN with the mediator
computer, which will then talk to the camera through its driver.

\subsection{Active Flexure Compensation}
\label{sec:afc}

Since SOXS will be installed at the Nasmyth focus of the NTT, during
an observation it will change its orientation with respect to the
gravity vector.  This will result in some flexures which might move
the target wwith respect to the spectrographs slit.  For this reason,
two piezo-actuated tip-tilt mirrors (TTM) are located in the common
path and will be used to correct for this effect.

The TTMs will be commanded by INS through the instrument PLC via
analog signals (one per axis). Since the TTMs are not a VLT standard
actuator, a ``special device'' will be developed. During observations,
this component will operate as a ``tracking axis'', updating in a loop
the position of the TTM depending on the rotator angle, via a look-up
table. Since the expected loop frequency is about 1~Hz, timing
constraints are not tight, so it has been decided to implement this
tracking loop entirely in the IWS.

\section{Secondary Guiding}
\label{sec:sg}

NTT is provided with a guide probe for automatic guiding. In order to
compensate for possible differential auto-guiding errors, the option
of performing secondary guiding was explored.

As a first step, we investigated the probability of finding suitable
secondary guiding stars in a typical SOXS observation.  To this end,
we simulated images of stars of several magnitudes (from $m_r = 16.5$
to $19.8$) by using the Advanced Exposure Time Calculator
(AETC)\cite{2016SPIE.9911E..2UU}. Simulations were repeated 100 times
per magnitude bin, using two different seeing values (0.8, typical
value expected at La Silla\cite{LaSillaSeeing}, and 1.6 arcsec, a bad
seeing scenario) and as they were observed with no filter or with the
SDSS $r$ filter.
The integration time was set to one second, in order to simulate a
secondary guiding loop with a frequency of 1~Hz.  All other values
used in the simulation (telescope and instrument transmission,
atmospheric extinction, efficiency of the detector, etc.) are
consistent with the NTT, the La Silla Observatory, and the ACQ
parameters.
Then, simulated star positions were fitted with a 2D Gaussian function,
and the error and the Root Mean Square (RMS) between the
expected and retrieved position of the centroid were
calculated.
Results show that, by using the ACQ camera in SOXS at the NTT, it is
possible to measure the center of a star brighter than $m_{r} = 19$
with an RMS of less than 0.1 arcsec, which is the specification for
target stability on SOXS spectrographs slit.

Furthermore, we simulated 10\,000 random telescope pointings on the
sky observable from the NTT. For each pointing we counted the number
of stars brighter than $m_R = 19.0$ in the ACQ FoV, available in the
General Star Catalog II (GSC2)\cite{2006IAUJD..13E..49S}.
Our results indicate that in more than $>95\%$ of pointing, at least
one suitable guide star can be found in the GSC2.

From this analysis, we can conclude that secondary guiding in SOXS is
feasible for the majority of observations.

\begin{figure} [t]
  \centering
  \includegraphics[width=0.59\textwidth]{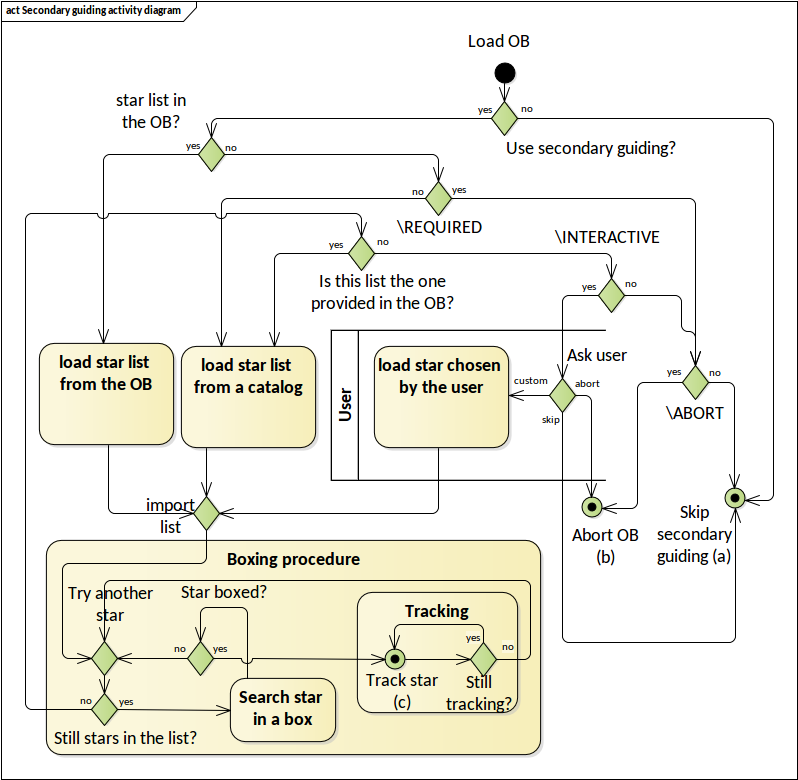}
  \caption[schematic-diagram]
  { \label{fig:secondary-guiding} SOXS Secondary Guiding (SG) sequence
    diagram.  Possible scenarios include running the Observation Block
    (OB) without the use of secondary guiding~(a), aborting of the
    OB~(b), or successfully track a star~(c).  Fallbacks are
    controlled by a series of flags in the OB: ``\texttt{REQUIRED}''
     determines if the guiding star must be provided in the OB;
     ``\texttt{ABORT}''  determines whether to skip SG or abort the
    observation;  ``\texttt{INTERACTIVE}''  determines if a user
    interaction is foreseen. }
\end{figure}

During spectroscopic observations, the central $15^{\prime\prime}$
(diameter) of the observed field will be transmitted to the
spectrographs, while the rest will be visible in the ACQ imager.  A
reference star will be selected in the ACQ field and used to check for
guiding errors. A secondary loop will run in SOXS IWS, at a lower
frequency with respect to the telescope auto-guiding, and will use the
reference star to measure error vectors and compute corrections, which
will be sent to the telescope as offsets.

Since SOXS is expected to perform observations of transients it is
desirable that its operations are as automatic as possible, with
little user interaction. This has been taken into account in designing
the secondary guiding star acquisition procedure shown in
Fig.~\ref{fig:secondary-guiding}.

If the observation foresees secondary guiding, a list of information
about suitable guiding stars in the target field (expected position,
expected brightness) may then be provided in the OB, and this
requirement can be either mandatory or not.  In a positive case, a
boxing procedure is called.  This procedure consists of a series of
operations to select a guiding star candidate from a list, search it
in a box, and track it.
If the guiding star list is not required nor available in the OB, or
the list provided in the OB does not help to successfully find an
object to track, then another star list from a catalog will be loaded.
The boxing procedure will then be called on this catalog list.
In case of failure of the boxing procedure, the user intervention may
be asked to skip the guiding, to abort the observation, or to pick
manually a star in the Acquisition Camera FoV, in order to repeat the
 procedure.

\section{Development Status}
\label{sec:dev}
 
The Software Development Process, mandated by ESO for a VLT Instrument
Software, foresees that no code is written until the Final Design
Review has been successfully granted.

In the SOXS case, however, we decided to proceed differently.  First,
because we were not bound to closely follow the VLT ``rules'', but
also because SOXS, being similar in design to X-Shooter, shares many
operational procedures with it, and so we could take the existing
X-Shooter INS as guidance.

\begin{figure} [t]
  \centering
  \includegraphics[width=0.66\textwidth]{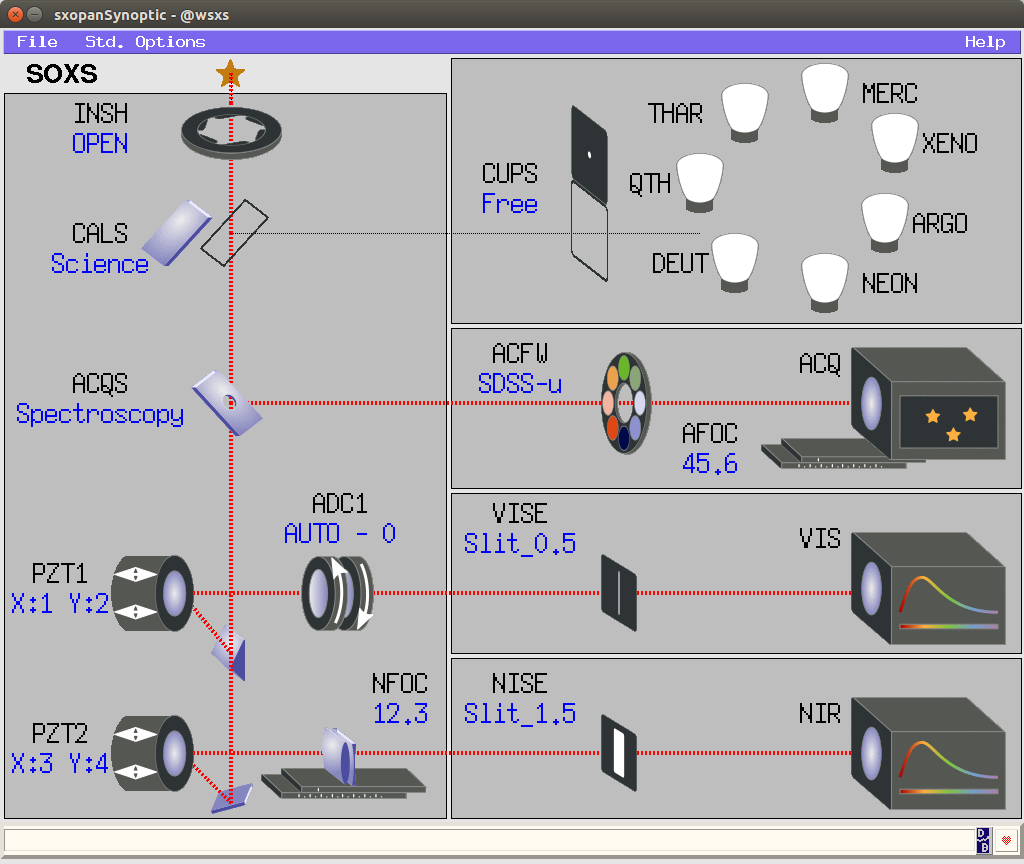}
  \caption[synoptic] { \label{fig:synoptic} SOXS Synoptic panel,
    showing a configuration in which light is directed to the
    acquisition camera. }
\end{figure}

Thus, during the design phase, we adapted the template Instrument
Software, provided by the VLTSW, to the SOXS case, and started
configuring standard (software) devices.  Dummy configurations were
prepared for the two science detectors (because the real ones are not
available yet). The template TDCS implementation was used for the
technical camera. This allowed us to run SOXS INS in full software
simulation and to start implementing and testing templates.
Procedures for starting-up and shutting-down the software have been
written, and a few user interfaces have been implemented.
Figure~\ref{fig:synoptic} shows the current implementation of the
Synoptic panel, which is intended to provide a quick graphical
representation of the current instrument setup.

Automatic tests have also been implemented, which excercise software
build from scratch, testing of the instrument functions, the detector
control software, the observation software and, finally, all
templates.
Currently, roughly the 50\% of observing templates, mainly concerning
calibration procedures, and the 20\% of maintenance templates have
been completed.

\section{Conclusion}
\label{sec:conc}

We presented the general software architecture of the SOXS instrument
for the NTT, a double spectrograph with an imaging and guiding camera,
and its peculiar features in terms of hardware (guiding camera and
camera connection) and software solutions (special devices, secondary
guiding).

The instrument software is based on the VLT Common Software for the
management of standard devices, which have already been configured in
simulation, and on custom software for the management of special
devices, i.e. the piezo-electric actuators for flexure compensation
and the piezo-mechanic slit exchanger for the near-infrared
spectrograph.

Spectrograph detectors are managed using ESO NGC controllers, while
the COTS imaging camera is managed through a custom component based on
TDCS.
We also presented a default (USB extender) and a fallback
(single-board computer) scenario for the connection of the imaging
camera to the instrument workstation.

Observation and Maintenance software have been completed mainly for
what concerns calibration (Dome Flats, Bias, Lamps, Motor tests), and
we estimate that about 40\% of SOXS templates are completed,
while development is ongoing for remaining templates.

Imaging camera is also responsible for secondary guiding.  We
presented a sequence of operations that minimize user intervention,
and we calculate that the probability to find at least two tracking
star candidates in the Field of View of the camera is $>95\%$.


\bibliography{ricci-soxs} 
\bibliographystyle{spiebib} 

\end{document}